# A Facility for Low-Radioactivity Underground Argon

Henning O. Back[1,*,†,‡], Walter Bonivento[2,§], Mark Boulay[3,‡,**], Eric Church[1,††], Steven R. Elliott[4,‡‡], Federico Gabriele[5,§], Cristiano Galbiati[6,7,§], Graham K. Giovanetti[8,§§], Christopher Jackson[1,††], Art McDonald[9,§§,**], Andrew Renshaw[10,‡], Roberto Santorelli[11,***], Kate Scholberg[12,††,†††], Marino Simeone[13,‡], Rex Tayloe[14,†††], Richard Van de Water[4,‡‡‡]

1. Pacific Northwest National Laboratory, Richland, Washington 99352, USA
2. INFN Cagliari, Cagliari 09042, Italy
3. Carleton University, Ottawa, Ontario K1S 5B6, Canada
4. Los Alamos National Laboratory, Los Alamos, New Mexico 87545, USA
5. INFN Laboratori Nazionali del Gran Sasso, Assergi (AQ) 67100, Italy
6. Princeton University, Princeton, NJ 08544, USA
7. Gran Sasso Science Institute, L'Aquila 67100, Italy
8. Williams College, Williamstown, MA 01267 USA
9. Queen's University, Kingston, ON K7L 3N6, Canada
10. University of Houston, Houston, Texas 77204, USA
11. CIEMAT, Madrid 28040, Spain
12. Duke University, Durham, North Carolina 27708, USA
13. Università degli Studi di Napoli "Federico II", Napoli 80125, Italy
14. Indiana University, Bloomington, Indiana 47405, USA

## Abstract/Executive summary

The DarkSide-50 experiment demonstrated the ability to extract and purify argon from deep underground sources and showed that the concentration of $^{39}$Ar in that argon was greatly reduced from the level found in argon derived from the atmosphere. That discovery broadened the physics reach of argon-based detector and created a demand for low-radioactivity underground argon (UAr) in high-energy physics, nuclear physics, and in environmental and allied sciences. The Global Argon Dark Matter Collaboration (GADMC) is preparing to produce UAr for DarkSide-20k, but a general UAr supply for the community does not exist. With the proper resources, those plants could be operated as a facility to supply UAr for most of the experiments after the DarkSide-20k production. However, if the current source becomes unavailable, or UAr masses greater than what is available from the current source is needed, then a new source must be found. To find a new source will require understanding the production of the radioactive argon isotopes underground in a gas field, and the ability to measure $^{37}$Ar, $^{39}$Ar, and $^{42}$Ar to ultra-low levels. The operation of a facility creates a need for ancillary systems to monitor for $^{37}$Ar, $^{39}$Ar, or $^{42}$Ar infiltration either directly or indirectly, which can also be used to vet the $^{37}$Ar, $^{39}$Ar, and $^{42}$Ar levels in a new UAr source, but requires the ability to separate UAr from the matrix well gas. Finding methods to work with industry to find gas streams enriched in UAr, or to commercialize a UAr facility, are highly desirable.

---

* Corresponding author. PNNL, 902 Battelle blvd., P.O. Box 999, MSIN J4-65, Richland, WA 99352 Email:henning.back@pnnl.gov
† Representing environment and applied sciences
‡ Representing the Urania portion of GADMC/Darkside-20k
§ Representing the Aria portion of GADMC/Darkside-20k
** Representing the DEAP-3600 collaboration
†† Representing the DUNE-like detector
‡‡ Representing the LEGEND collaboration
§§ Representing the GADMC/Darkside-20k collaboration
*** Representing the DART portion of GADMC/Darkside-20k
††† Representing the COHERENT collaboration
‡‡‡ Representing the Coherent Captain-Mills (CCM) collaboration

# Introduction

Argon derived from the atmosphere, which is predominately stable $^{40}Ar$ [1], contains cosmogenically produced long-lived radioactive isotopes of argon— $^{42}Ar$, $^{39}Ar$, and $^{37}Ar$ [2]. These isotopes are at high enough concentrations in the atmosphere to be a significant background for low-background argon-based radiation detectors, and because all commercial argon is produced from air, these argon radionuclides can represent irreducible backgrounds. However, after a 5-year campaign of extracting and purifying argon from deep $CO_2$ wells in Southwestern Colorado in the United States, the DarkSide-50 dark matter search experiment demonstrated that this unique argon contained an $^{39}Ar$ concentration 0.073% of that in the atmosphere, or approximately 0.73 mBq/kg$_{Ar}$ [3]. Cosmogenically produced $^{37}Ar$ was also detected in the early running of the DarkSide-50 detector [3], but $^{42}Ar$ was not observed. This demonstrated that large-mass quantities of low-radioactivity underground argon can be obtained and has sparked a global interest in a sustained supply of low-radioactivity underground argon to meet the needs of a broad range of disciplines, from nuclear and particle physics to environmental studies and national security.

The broader availability of low-radioactivity argon and the challenges associated with finding new sources, the current and future production of underground argon, and the characterization of underground argon have become important topics associated with the growing need. As an example, measuring the concentration of $^{39}Ar$ in underground argon is critical to both searching for new sources and as quality assurance and quality control during production. Additionally, the determination of argon radionuclide production underground and in the atmosphere is relevant to both new underground argon source characterization and in understanding cosmogenic activation when the underground argon resides close to, or on the surface of, the Earth. Both issues are also relevant in the context of groundwater dating using $^{39}Ar$. Additionally, isotope separation may be required to achieve the level of argon-radioisotope depletion needed for some applications.

# The need for low-radioactivity underground argon (UAr)

The largest needs for low-radioactivity underground argon are in the fundamental nuclear and particle physics fields. The DarkSide experiments have been driving the demand and production for low-radioactivity underground argon [3], but with that success the demands for UAr have risen. Beyond WIMP dark matter detection, the physics that is more easily reached by the availability of low-radioactivity underground argon includes: neutrinoless double-beta decay, by eliminating $^{42}Ar$ and $^{39}Ar$ in the argon that surrounds the germanium crystals of the LEGEND experiment; measuring low-energy neutrinos in a DUNE-like detector, by reducing the $^{39}Ar$ beta rate and also the higher-energy beta from $^{42}K$ (the daughter of $^{42}Ar$); and coherent elastic neutrino-nucleus scattering within the series of COHERENT experiments, by increasing live-time by reducing $^{39}Ar$ decays.

The potential needs for low-radioactivity underground argon span from tens of kilograms for the COHERENT experiment to tens of kilotonnes for a DUNE-like module to a long-term, enduring supply for low-level radiation detection for environmental science (10-100 kg annually). Likewise, the requirements for $^{39}Ar$ reduction from the atmospheric concentration span several orders of magnitude, from a factor of 10 reduction for COHERENT to more than a factor of 1000 for future argon-based dark matter searches. The specific needs for low-radioactivity underground argon are reviewed for five distinct uses.

## The Global Argon Dark Matter Collaboration

DarkSide-20k is a 20-tonne fiducial volume dual-phase TPC to be operated at LNGS with 100 tonnes of underground argon (UAr), designed to collect an exposure of 100 tonne×years, completely free of neutron-induced nuclear recoil backgrounds and all electron recoil backgrounds [4]. DarkSide-20k will have sensitivity to WIMP-nucleon spin-independent cross sections of $1.2 \times 10^{-47} cm^2$ for WIMPs of 1 TeV/c$^2$ mass following a 5-year run. An extended 10-year run could produce an exposure of 200 tonne×years, with sensitivity to a cross section of $7.4 \times 10^{-48}$ cm$^2$, for the same WIMP mass. DarkSide-20k will explore the WIMP-nucleon cross-section down to the edge of the "neutrino floor," where coherent neutrino-nucleus scattering from environmental neutrinos induce nuclear recoils in the detector. At the same time, the Global Argon Dark Matter Collaboration (GADMC) will pursue the search for low-mass WIMPs with a 1-tonne detector, DarkSide-LowMass, which also takes aim at the "neutrino floor" for masses below 8 GeV/c$^2$. The level of $^{39}$Ar depletion directly impacts the sensitivity of DarkSide-LowMass, therefore it is planned to be run with UAr processed for further reduction of $^{39}$Ar with the Aria plant in Sardinia, the tallest cryogenic distillation plant ever constructed [5].

Plans are being developed within the GADMC for an experiment beyond DarkSide-20k for further WIMP reach, Argo. Argo will be a multi-hundred tonne detector that will approach the neutrino floor following a 1,000 tonne-year exposure. The Argo will require upwards of 500 tonnes of low-radioactivity underground argon [6].

## COHERENT NEUTRINO SCATTERING

Liquid argon experiments searching for low-energy beam-related processes, such as coherent elastic neutrino nucleus scattering (CEvNS) [7, 8] and the production of dark matter particles in a proton beam [9] could benefit from the reduction of $^{39}$Ar provided with the use of UAr. Specific examples are the liquid argon detectors employed by COHERENT at the ORNL SNS [10] and by Coherent Captain-Mills (CCM) at the LANL Lujan center [11, 12]. In both of these scintillation-only detectors, both electron/gamma identification and the use of a pulsed beam greatly reduces the $^{39}$Ar background, but it is still the dominant steady-state background [13]. The use of UAr with a $^{39}$Ar suppression factor of O(x100-x1000) would substantially improve the statistical precision of the resulting measurements [14].

## DUNE-LIKE DETECTOR

A multi-kiloton liquid argon time projection chamber would have unique sensitivity to the electron neutrino component of a core-collapse supernova burst, via the dominant charged-current (CC) interaction $\nu_e + ^{40}\text{Ar} \rightarrow e^- + ^{40}\text{K}^*$ [15]. The dominant radiological background is $^{39}$Ar, the presence of which degrades event reconstruction quality and presents a challenge for DAQ and trigger design. For example, even though argon allows to perform e$^-$/γ background removal via pulse shape discrimination (PSD), for a WIMP dark matter search, a few kilo-tonnes argon volume will create a pile-up problem, without the required low level of $^{39}$Ar suppression that UAr provides. The background is furthermore a serious impediment to potential detection of the "CEvNS glow" of a supernova burst [16] and will impede other ambitious attempts to extract information from very low energy physics signals [17]. Obtaining quantities of UAr sufficient to fill a DUNE-like detector will be challenging, but there have been positive discussions with commercial underground gas producers indicating that on the scale of 10s of kilo-tonnes, UAr might be produced in an acceptable time scale (approx. 5 years) at a roughly estimated cost of only three times commercial argon.

## ENVIRONMENTAL ASSAY

Radiometric age dating using $^{39}$Ar covers an intermediate age range not achievable with other naturally occurring radioisotopes [18] and is used in fields such as ground water age dating to understand the recharge rate of the California Central Valley Aquifer system [19], ocean ventilation [20], and others. A handful of groups around the globe have developed technologies for measuring trace amounts of $^{39}$Ar [19, 21, 22, 23]. These technologies will benefit greatly from a supply of argon free of $^{39}$Ar for detection limit determination and to avoid the introduction of modern argon (which is $^{39}$Ar rich) into the detection systems.

It has been estimated that access to 100 kg/year for many years (i.e., decades) would satisfy both current and expanding needs in this area. Although this could be met by stockpiling from the UAr production for fundamental physics experiments, a truly enduring supply would be beneficial as underground storage becomes necessary for the long-term, and institutional knowledge about how the stockpile was obtained may be lost over time. It is not likely that an $^{39}$Ar concentration better than that attained in the DarkSide-50 experiment would be required.

## LEGEND

The Large Enriched Germanium Experiment for Neutrinoless double beta Decay (LEGEND1000) is proposed to search for neutrinoless double beta decay (0νββ) of $^{76}$Ge. The experiment will probe the 0νββ decay of $^{76}$Ge with a 3σ discovery potential of a half-life > $10^{28}$ yr corresponding to a limit on the effective neutrino mass of < 17 meV. The Ge detectors will be distributed among four 250-kg modules to allow commissioning of the array in stages and independent operation. In each module, the detector strings are immersed within a liquid argon active shield, sourced from radiopure UAr. It is expected that $^{42}$Ar will be greatly reduced within underground Ar, as is the case for $^{39}$Ar. Hence, this UAr will provide direct reduction of background from the decay of $^{42}$K, a daughter of $^{42}$Ar decay.

# Current UAr production and future facility

The current large-scale production of UAr is being pursued by the Global Argon Dak Matter Collaboration (GADMC) focused on production for the DarkSide-20k detector. To create UAr of sufficient quality for the DarkSide-20k experiment, argon is first extracted from $CO_2$ in southwestern Colorado that is used for enhanced oil recovery [24]. The crude UAr is then shipped to Sardinia where it is purified through cryogenic distillation. The plants that are used for this UAr production currently fall under the purview of the GADMC, but when the UAr for DarkSide-20k has been obtained, then those plants could become part of a UAr facility to provide UAr for a number of needs; the primary needs being those listed above.

Urania is the plant that will be located outside of Cortez, Colorado to produce the crude UAr. It will accept $CO_2$ with 400 ppm of UAr and will produce 99.9% pure UAr at a rate of 330 kg/day. The plant is largely based on cryogenic distillation, where a distillation column is used to remove bulk $CO_2$ and then a pressure swing adsorption unit is used to fully remove all $CO_2$. Then the gas passes through 2 more distillation columns, first to remove methane and then to remove nitrogen. The plant will be installed in CY23 with full operations in CY24 and the final UAr extracted for DarkSide-20k in mid CY25. [4]

Aria is a 350 m tall cryogenic distillation column, the tallest distillation column in the world, capable of isotopic enrichment [25]. The small difference in the vapor pressure of argon isotopes [26] allow for the separation of argon isotopes by distillation. Operating in a mine shaft on the island of Sardinia in Italy, Aria will be able to further reduce the concentration of $^{39}$Ar by a factor of 10 per pass and at a rate of several kg/day. Beyond argon isotopic enrichment, the column has commercial applications in the production of isotopes for nuclear energy and medicine. For DarkSide-20k, however, Aria will not be used to reduce $^{39}$Ar, but rather to chemically purify the crude UAr from Urania (99.9% pure) to produce detector-grade UAr. For this chemical purification Aria will produce on the order of 1000 kg/day of purified UAr.

# Quality Control and Quality Assurance

For any source of UAr, it is crucial to eliminate, or to minimize to acceptable levels, the infiltration or creation of $^{37}$Ar, $^{39}$Ar, and/or $^{42}$Ar (radioargon) in the UAr during production and storage, either due to poisoning the UAr with atmospheric argon or the production of radioargon by cosmic rays. The allowable radioargon content is experiment specific, and a general use UAr productoin facility should be designed to meet the most stringent experimental requirement, albeit with a lower limit imposed by the intrinsic purity of the crude UAr containing gas. Once the UAr has reached the surface of the earth, cosmogenic activation becomes the dominant production mechanism for radioargon. The level of effort required to transport and shield the UAr from cosmic rays will be driven by the particular use case.

### Direct radioargon level monitoring

The levels of radioargon in the UAr and levels required by experiments are far lower than what is detectable in most readily available assay capabilities. Within the age-dating communities, there are 2 radiometric laboratories in the world that measure $^{39}$Ar and $^{37}$Ar using gas proportional counting, The University of Bern in Switzerland and Pacific Northwest National in the United States [27, 19]. An emerging technology is the use of atom trap trace analysis, which has been demonstrated in measuring $^{39}$Ar for understanding ocean ventilation [20]. These technologies are able to reach $^{39}$Ar levels at approximately 5% of the atmospheric concentration, or about 50 mBq/kg, which is only sufficient if there is a gross infiltration of atmospheric argon into the UAr process stream. However, the sample sizes are from 10cc to 1 liter of argon at STP and may be of benefit for rapid sampling. A dedicated detector for measuring ultra-low levels of $^{39}$Ar in argon, DART, has been built within the former ARDM detector [28]. DART is a small (~1 L) chamber that will measure the depletion factor of $^{39}$Ar in UAr to levels a factor of 1000 below the atmospheric concentration (~1mBq/kg). The sample size required for DART is about 2.2 kg [28]. The radiometric technologies can also be used for $^{37}$Ar, but, although it has been measured in the atmosphere, there is no method for routine assay of argon for $^{42}$Ar.

Each of these $^{39}$Ar assay technologies are fixed laboratory capabilities and require time for analysis. Therefore, their utility in monitoring $^{39}$Ar levels in UAr production are limited to a batch mode check, where UAr samples are taken at the production plant and sent to the facility. This calls into question the method of transportation and if it is permissible to expose the UAr to greater cosmic ray flux in flight to an assay facility, or if it must remain on the ground/sea, which will increase sample travel time. Assay systems may be required to be at least within the same continent as the UAr production facility to minimize cosmic ray exposure.

### Indirect radioargon monitoring

An indirect method for monitoring radioargon contamination in the UAr is to search for air infiltration during production. The 0.934% argon concentration in atmospheric argon makes monitoring for air infiltration increasingly difficult with lower UAr concentrations in the crude gas. The concentration of UAr in the $CO_2$ stream that the DarkSide-50 UAr was derived from has a concentration of about 400 ppm [29]. At this level, the amount of $^{39}$Ar found in the DarkSide-50 UAr would be equivalent to approximately 30 ppm of air in the $CO_2$ stream. When the well gas also contains air constituents (e.g. nitrogen), the monitoring of air infiltration is further complicated. If oxygen is used as a tracer for air infiltration, then to measure 30 ppm of air requires a 6 ppm sensitivity to oxygen, but sensitivities as low as 1 ppb may be needed for well gas with low concentrations of UAr and/or driven by experimental requirements. If the well gas contains nitrogen and oxygen, then other air constituents might be used, which will require instrumentation with even greater sensitivities to those gases. This mode of monitoring is UAr source specific, because it depends on the constituents of the crude UAr containing well gas.

An alternate solution that has been developed is to use the ratio of stable argon isotopes [30]. In underground argon the abundance of $^{36}$Ar and $^{38}$Ar are lower than in argon derived from the atmosphere [31]. Therefore, the $^{36}$Ar:$^{40}$Ar ratio can be used to monitor the final UAr product for any air infiltration in the entire UAr production process [30]. This mode of monitoring will depend on the how sensitively the ratio must be monitored to detect an air infiltration of concern.

## A facility beyond Urania/Aria

For UAr needs that are beyond the means of Urania and Aria, whether that be because quantities of UAr are needed beyond what the Urania plant can provide or that Urania/Aria have reached their end of life, a new source will need to be found. For very large-scale production, for example what may be needed for a DUNE-like experiment, to make UAr production the most cost efficient would require finding sources that either have high argon concentrations, or where the UAr is chemically enriched to higher concentrations. For example, helium is a gas that is extracted and purified from natural underground gases [32]. If that underground gas also contains UAr, then it is highly likely that there is a process stream with enriched UAr. At PNNL, discussions with major gas producers have identified exactly such a stream, where there is enriched UAr and would be easier to process than directly from the crude underground gas. The commercial producer has estimated that for a 10+ kilo-tonne-scale production the UAr may cost a factor three more than regular commercially available argon. Such a future UAr facility may simply become a parasitic operation to the major underground gas production. Ideally, UAr production would be carried out by a commercial producer, for which the Urania plant might still be used in whole or in part to produce UAr from the new enriched source.

To find a new source, the level of radioargon underground needs to be predicted. Estimating $^{39}$Ar production in underground minerals has been carried out [33], but how it enters a gas field needs to be understood. The radioargon levels will need to be verified and the detection systems that are used for QA/QC can be used for a new UAr source characterization, but methods for separating UAr from the matrix it is found in are needed. This becomes more challenging the more UAr that is needed (~2kg for DART). Systems exist for separating argon from air [34], but new systems will be needed the more the matrix deviates from air (e.g., $CO_2$ wells).

## Conclusion

The work of the Darkside collaboration to develop a source of UAr with Urania/Aria has the potential to benefit a host of other low-background experiments with modest incremental investment. With sufficient resources, operating the Urania and Aria infrastructure beyond the production period required for DarkSide-20k could meet the UAr demand for most of the foreseen future experimental needs. However, if the UAr source in southwestern Colorado becomes unusable in the future, then a new source would need to be found and Urania may be incorporated into that new facility. For the production of very large masses of UAr, e.g., for a DUNE-like detector, it is envisioned that a commercial source would be required, where UAr is found enriched in part of the process stream for other gas separations. For this effort to be successful, an array of ancillary facilities and technologies must be developed, primarily focused around quality control and assurance, and these technologies may be used to search for new UAr sources.